\documentclass[final,5p,authoryear]{elsarticle}

\usepackage{amssymb}
\usepackage{acronym}
\usepackage{stfloats}
\usepackage[table,xcdraw]{xcolor}
\usepackage{amsmath,amsfonts}

\journal{arXiv}

\begin{document}

\begin{frontmatter}



\title{Fuel Saving Effect and Performance of Velocity Control for Modern Combustion-Powered Scooters}
\author[1,2]{Jannis Kreß \corref{cor1}}
\ead{jannis.kress@fb2.fra-uas.de}
\author[1]{Jens Rau}
\author[1]{Hektor Hebert}
\author[2]{Fernando Perez-Peña} 
\author[1]{Karsten Schmidt} 
\author[2]{Arturo Morgado-Estévez}

\cortext[cor1]{Corresponding author.}
\newcommand{\abbreviations}[1]{%
  \nonumnote{\textit{Abbreviations:\enspace}#1}}

\affiliation[1]{organization={Department of Computing and Engineering, Frankfurt University of Applied Sciences},
             city={Frankfurt},
             postcode={60318},
             state={Hessen},
             country={Germany}}
\affiliation[2]{organization={Department of Automation, Electronics and Computing Architecture and Networks, University of Cadiz},
             city={Puerto Real},
             postcode={11519},
             state={Andalusia},
             country={Spain}}

\begin{abstract}
This paper investigates the performance and fuel-saving effect of a velocity control algorithm on modern 50 cc scooters (Euro 5). The European Parliament has adopted major CO$_2$ emission reductions by 2030. But modern combustion-powered scooters are inefficiently restricted and emit unnecessary amounts of CO$_2$. Replacing the original restriction method with the system presented in this paper, the engine's operating point is being improved significantly. Therefore, a Throttle-by-Wire-System senses the rider's throttle command and manipulates the throttle valve. A redundant wheel speed sensor measures the precise vehicle velocity using the magneto-resistive principle. The entire system is managed by a central ECU, executing the actual velocity control, fail-safe functions, power supply and handling inputs/outputs. For velocity control, an adaptive PI-controller has been simulated, virtually tuned and implemented, limiting the max. velocity regulated by legal constraints (45 km/h). In this way, the environmentally harmful restrictors used today can be bypassed. By implementing a human-machine interface, including a virtual dashboard, the system is capable of interfacing with the rider. For evaluation purposes a measurement box has been developed, logging vehicle orientation, system/control variables and engine parameters. A \textit{Peugeot Kisbee 50 4T} (Euro 5) is serving as test vehicle. Finally, the system has been evaluated regarding performance and fuel efficiency both through simulation and road testing. Fuel savings of 13.6 \% in real-world test scenarios were achieved while maintaining vehicle performance.
\end{abstract}



\begin{keyword}
Velocity control \sep Throttle-by-Wire \sep Fuel saving \sep Motorcycle powertrain \sep Alternative restricting
\end{keyword}

\end{frontmatter}

\newacro{CVT}{Continuously Variable Transmission}
\newacro{EC}{Engine Controller}
\newacro{ECU}{Electronic Control Unit}
\newacro{HMI}{Human Machine Interface}
\newacro{MB}{Measurement Box}
\newacro{MECU}{Main ECU}
\newacro{TbWS}{Throttle-by-Wire-System}
\newacro{TDC}{Top Dead Center}
\newacro{TPS}{Throttle Position Sensor}
\newacro{TSS}{Transmission Speed Sensor}
\newacro{TVA}{Throttle Valve Actuator}
\newacro{VC}{Velocity Control}
\newacro{WSS}{Wheel Speed Sensor}

\let\UrlSpecialsOld\UrlSpecials
\def\UrlSpecials{\UrlSpecialsOld\do\/{\Url@slash}\do\_{\Url@underscore}}%
\def\Url@slash{\@ifnextchar/{\kern-.11em\mathchar47\kern-.2em}%
    {\kern-.0em\mathchar47\kern-.08em\penalty\UrlBigBreakPenalty}}
\def\Url@underscore{\nfss@text{\leavevmode \kern.06em\vbox{\hrule\@width.3em}}}

\section{Introduction}
According to the climate targets of the European Climate Change Act, at least 55 \% of greenhouse gases are to be saved by 2030 compared to 1990 levels \citep{EUcca}. Considering the average distance traveled per inhabitant in the EU, the values vary between 5 and 20 km a day. Depending on the country of origin, 57 \% to 81 \% is covered by car. Most often, the distance traveled by two-wheelers corresponds to less than 1 \% \citep{EUpms}. However, for urban mobility, scooters are an excellent alternative. The potential for saving fuel and thus minimizing CO$_2$ emissions is enormous and can make a significant contribution to EU climate targets. Harmful two-stroke engines are banned and the Euro 5 standard also applies to 50 cc scooters \citep{EURO5}. Comparing modern four-stroke combustion and electrically powered scooters of Peugeot, Piaggio and Vespa, an average price for electric models of almost 160 \% can be noticed. The difference in range is even more striking, as the combustion engine exceeds that of the electric drive by a factor of 6. Users require longer ranges because recharging is often difficult in cities.

To fulfill Euro 5 requirements, modern combustion-powered scooters are equipped with gasoline injection and a catalytic converter \citep{Peugeot, Kymco}. To ensure optimal operation of the catalyst, the air/fuel mixture ($\lambda$) must be approx. 1 \citep{lambda}. Legal requirements state that 50 cc scooters cannot exceed a maximum speed of 45 km/h. By delaying the ignition timing, the engine is restricted in power, since reducing the fuel injected would cause $\lambda$ to become larger 1 (lean mixture). This can be solved by deactivating the original restriction of the test vehicle and applying a \ac{VC} algorithm. The basis is provided by a low-cost \ac{TbWS} \citep{TbWS_JK}, which enables electrical control of the throttle valve. For this purpose, the throttle position is sensed and the throttle valve is manipulated by an actuator. As a consequence, instead of the ignition timing, the airflow is regulated by means of the throttle valve position, which also results in a reduction of the fuel injection volume. This system can contribute substantially to reaching EU climate targets. To the best of our knowledge, this is the first approach of a \ac{VC} application on a modern four-stroke 50 cc (Euro 5) scooter as an alternative restriction. The \ac{VC} system investigated here is intended to optimize fuel economy and improve exhaust emissions. 
\newpage
The following novelties will make a significant scientific contribution in this class of vehicles:

\begin{itemize}
    \item By suppressing ignition timing manipulation, the efficiency is increased and the throttle valve opening can be reduced by almost 50 \% at top speed. 
    \item The controlled air supply minimizes the amount of fuel injected in line with the optimum stoichiometric ratio by up to approx. 20 \% at top speed.
    \item Based on the frequently maxed-out vehicle velocity, this alternative restriction significantly lowers fuel consumption/CO$_2$ emissions by 14 \%.
    \item Despite the minimized injection quantity, consistent performance can be demonstrated because the amount of energy supplied is used more effectively.
\end{itemize}

\subsection{Background on restricting methods}
\begin{figure}[!h]
\centering
\includegraphics[width=8.5cm]{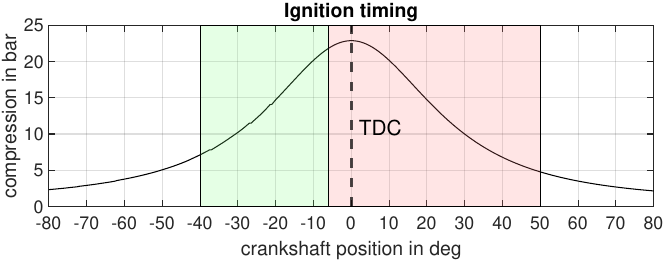}
\caption{Ignition timing}
\label{IgnT}
\end{figure}

Scooters are defined by EU law. In addition to the max. speed of 45 km/h and the max. engine capacity of 50 cc, the power is also limited to 4 kW, regardless of the type of drive \citep{Def_scooter}. In most cases, vehicle performance is sufficient to reach higher speeds, so the surplus power above 45 km/h must be compensated. The surplus is still needed for heavy loads or inclines. In the past, mechanical restrictors were often used, which manipulated the flow behavior either through throttle orifices in the air intake or exhaust \citep{R1}. Limitation of the \ac{CVT} was particularly common, resulting in an unnecessarily high engine speed. Electronic throttling was opened with the arrival of electronic ignition. When the top speed was reached, the ignition was interrupted, resulting in the emission of unburned fuel. Modern powertrains have further capabilities to influence engine control due to gasoline injection. Injection volume and ignition timing are decisive here. Reducing the fuel supplied would be obvious, but leads to high combustion temperatures and interferes with the function of the catalytic converter \citep{R2}. As a consequence, engine damage and unacceptable exhaust emissions would be expected. Typically, ignition occurs 6° to 40° before the \ac{TDC}, depending on the engine speed. For restriction, the ignition timing is delayed towards the \ac{TDC} or even behind as shown in  Figure \ref{IgnT} (efficient left, restricted right). 

To achieve optimum efficiency, the maximum pressure must be reached shortly after \ac{TDC}. This ensures that the max. combustion chamber pressure can act on the piston from the start of expansion. By retarding the ignition timing, expansion is already taking place during combustion, thus reducing both the peak pressure and the usable expansion distance. The efficiency drops \citep{IgnT}.

\subsection{Background on \acp{TbWS} \& velocity control }
\acp{TbWS} have been researched and integrated industrially to motorcycles \citep{TbWS_SportMotorbike, TbWS_BMW1, TbWS_BMW2}. A \ac{TbWS} has also been applied successfully to small-volume engines, coming with the minimum number of low-cost components, combined in the throttle body. The system is based on a throttle servo, throttle position sensor, throttle demand and speed sensor. To limit and adapt the engine speed in various load situations, the speed can be adjusted by an integral electronic control and resonant frequencies can be avoided. Besides the performance limitation, the system is capable of managing speed and load reductions during overheating. For cold starts, the idle speed can be adjusted on the outside temperature \citep{EngineManagement}. Also, an electronic throttle control for a 50 cc two-stroke scooter application has been developed, improving fuel economy, idle-stability and implementing an electronic vehicle velocity control. By use of an electronic throttle control, the conventional \ac{CVT} restriction can be replaced, which leads to fuel savings of 22 \% caused by the lowered engine speed. Additionally, the system allows controlling the idle engine speed after the start-up. The system is based on a carburetor-driven two-stroke scooter with a power of 4.2 kW and a twin variator pulley \ac{CVT} \citep{ETC_Peugeot}. Due to the mentioned powertrain, the system is not compliant with the Euro 5 standard. Nowadays systems use an originally optimal-designed \ac{CVT}. In addition, modern air-cooled four-strokes have a power output of max. 2.5 kW due to the significantly lower power density. The potential within power surplus and transmission thus no longer exists. In addition, the fuel consumption of modern non-optimized scooter drives is considerably lower than that of an optimized two-stroke \citep{2vs4-Stroke}. 

\section{System development}
\begin{figure}[!h]
\centering
\includegraphics[width=8.1cm]{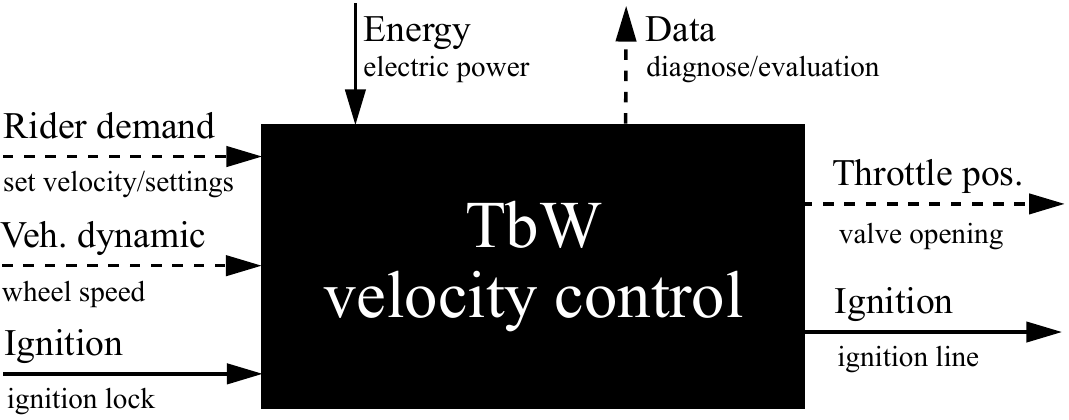}
\caption{Blackbox schematic}
\label{Blackbox}
\end{figure}

The TbW-based \ac{VC} uses the set velocity (throttle grip position) specified by the rider as the command variable/ input. A \ac{WSS} measures the vehicle's dynamics (velocity), used as control reference variable. Driving mode and settings serve as additional rider input. The desired vehicle velocity is manipulated through an electronically set throttle valve position (output). For safety reasons, the system is capable of interrupting the engine's ignition line. Lastly, the system requires power supply, outputs diagnose data and rider-intended system information for evaluation purposes. Figure \ref{Blackbox} shows a general black box schematic.\newpage

\begin{figure*}[!b]
\centering
\includegraphics[width=18.3cm]
{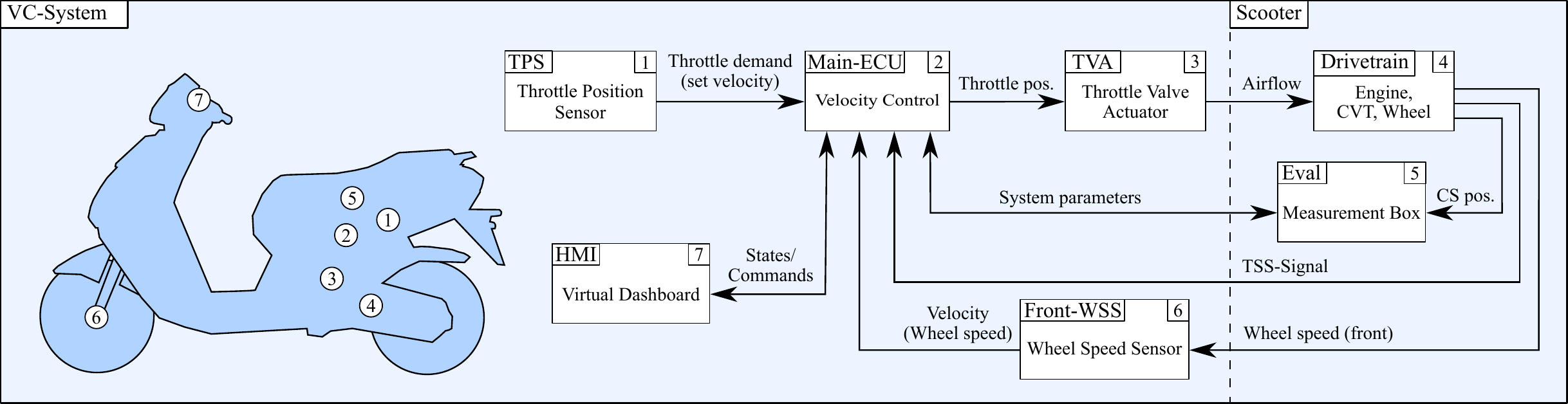}
\caption{System layout}
\label{SysLayout}
\end{figure*}

Several bus-connected subsystems are required to implement \ac{VC} and efficient restriction. The sensing of the rider's demand (throttle grip position) and the vehicle velocity must be achieved. To be able to influence the scooter's longitudinal driving dynamics, the throttle valve position must be controlled electronically. In parallel, emulation of the \ac{EC} is essential to prevent ignition timing manipulation. The remaining drive-train, consisting of the engine, gearbox and rear wheel, retains its original condition. A central \ac{ECU} processes input variables, performs \ac{VC}, provides necessary data on the bus and ensures safe vehicle operation. Furthermore, the \ac{ECU} serves as powerbox, providing electric energy to all subsystems. The overall system shall be designed in such way, as to allow switching between restrictions by ignition timing and throttle valve manipulation. By means of a virtual dashboard, system quantities and an eco-score are to be displayed to the rider. For evaluation, this \ac{HMI} will be used to switch between restrictions and to start measurements. These are to be recorded by a measurement box accessing the vehicle bus. Now that the foundation has been laid for assistance systems, cruise control is to be implemented. This section presents the architecture developed, including the sensors and systems needed. Figure \ref{SysLayout} gives the system layout which has been outlined. 

\subsection{Throttle-by-Wire-System}
The basis for the \ac{VC} implementation is provided by the previously developed low-cost \ac{TbWS}. A redundant, contactless AMR sensor is used to sense the throttle position with a measuring accuracy of 99.84 \%. The measured and plausibility-checked position is sent via CAN by the \ac{TPS}. To enable the system's influence onto longitudinal vehicle dynamics, a \ac{TVA} was developed. By determining the position using a Hall sensor in combination with a stepper motor, the throttle valve can be precisely position-controlled. This allows positioning with an accuracy of 99.63 \% within max. 60 ms. \ac{TPS} and \ac{TVA} make use of embedded fail-safe functions, monitoring measurement and position quality. Design, function, reliability and performance have already been evaluated \citep{TbWS_JK}. 

\subsection{Wheel speed sensor}
\begin{figure}[!h]
\centering
\includegraphics[width=8cm]{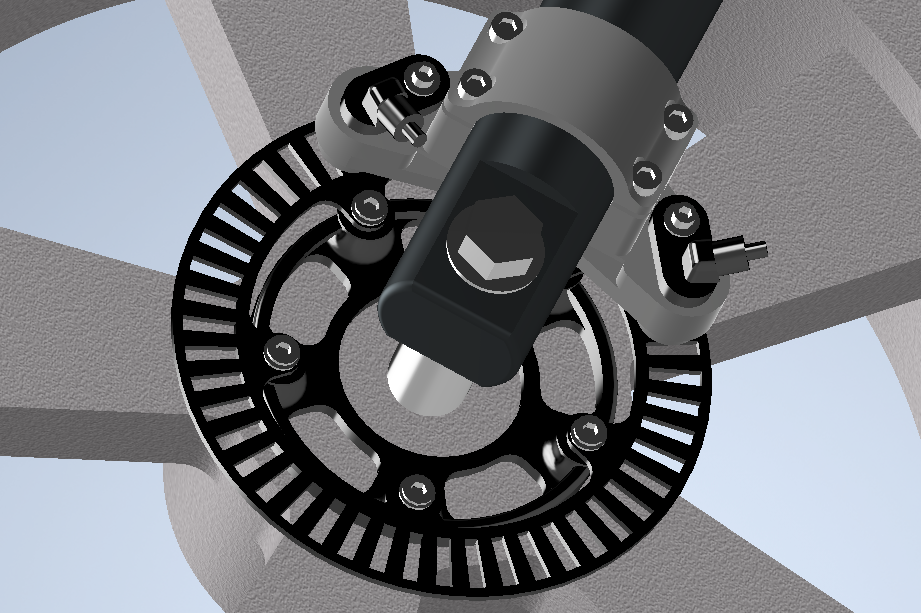}
\caption{\ac{WSS} adaption}
\label{WSS_adap}
\end{figure}

The scooter's velocity is required as a control variable for \ac{VC}. Since 50 cc scooters are not legally obligated to be equipped with an anti-lock braking system, they do not have \acp{WSS} \citep{ABS_law}. In the test vehicle, the rear wheel speed is only measured with a passive inductive \ac{TSS} on the clutch bell for restriction purposes. Due to its design, it is not suitable for measuring low speeds. Therefore, a redundant \ac{WSS} was developed, which enables the precise sensing of the wheel speed. Two automotive magneto-resistive \acp{WSS} independently measure the motion of a 48-step encoder disc by determining the width of successive pulses. A sampling rate of 1 MHz results in a resolution of 0.02 km/h. The \ac{WSS} assembly was tested within a hardware-in-the-loop environment \citep{WSS_JK}. The assembly is to be integrated, whereby a constant distance ($<$0.5 mm) between sensors and the encoder disc is to be ensured. Preferably, the controlled rear wheel speed is sensed to intervene in case of wheel slip. Adaptations were developed in several iterations. Unlike chain-driven motorcycles, there is no decoupling between the driven wheel and the engine/gearbox. Vibrations from strong low to high frequencies are directly transferred to the \ac{WSS} resulting in extreme measurement inaccuracies. Decoupling, damping and software filters for signal processing could not achieve sufficient signal quality. For that reason, the \ac{WSS} is adapted to the front wheel. In addition, the \ac{TSS} is used to detect wheel slip/lock at the rear. With a bracket, both sensors are precisely placed, as shown in Figure \ref{WSS_adap}.\newpage

\vspace{-0.2cm}
\subsection{Main \ac{ECU}}
The \ac{MECU} (\textit{STM32G431KB}) \citep{STM32} is intended to act as a higher-level system master. On the input side, this includes processing the wheel speed, receiving the \ac{TPS} measured value and the rider command from the \ac{HMI}. On the output side, the \ac{TVA} position, \ac{HMI} system states, emulated \ac{TSS} signal and evaluation data must be transmitted. At the hardware level, the \ac{MECU} must supply the entire system with power, provide the measurement electronics of the \acp{WSS} and enable the amplification of the emulated speed signal. At software level, the \ac{VC} needs to be implemented, adjusting the vehicle's velocity to the rider's demand. Here, the continuous functional safety of the overall system must be ensured by means of appropriate fail-safe functions. \\

\textbf{Power box:} The \ac{MECU} provides individually fused voltages to all subsystems. Especially the \ac{TVA} motor is to be supplied with an independent power line to prevent system errors in case of damage. Three 12 V DC/DC step-down converters provide stabilized voltages of 3.3 V (\ac{MECU} internal), 6 V (\ac{TVA} motor) and 7.5 V (common system voltage). In addition, the 12 V battery voltage is also provided. A wiring harness originates from the \ac{MECU} that connects all subsystems, consisting of 6 wires: GND, VCC$_{step}$, VCC$_{sys}$, VCC$_{bat}$, CAN$_{high}$ and $CAN_{low}$.\\

\textbf{\ac{WSS} processing:} A square-wave signal voltage is generated via a measuring resistor in the supply line of the \acp{WSS}. The signal conversion must be performed by the \ac{MECU} for the \ac{WSS} (front) and \ac{TSS} (rear). The \ac{WSS} signal acquisition can be done by external interrupts due to the constant measuring voltage. For processing the \ac{TSS}, a comparator is used to digitize the varying voltage of the inductive sensor. Afterward, the given pulse widths are measured by an input capture timer. The stored pulse widths (\textit{$T_{wss}$}) are smoothed by a moving average filter with the array size of (\textit{$s_{A}$}) before velocity conversion (\textit{$v_{scooter}$}). For the determination, the wheel circumference (\textit{$D_{wheel}$}) and the resolution of the encoder disc (\textit{$n_{steps}$}) are required. Equation (\ref{eq.1}) transfers the pulse widths into a velocity. In order to prevent dead time behavior, the \ac{WSS}'s array size (\textit{$s_{A}$}) adapts to vehicle velocity. Due to the redundant signals, errors in velocity measurement can be detected.

\begin{equation}
    v_{scooter}=\frac{d_{step}}{t_{step}}=\frac{D_{wheel}}{n_{steps}}\cdot{\frac{s_{A}}{\sum_0^{s_{A}}T_{wss}}}
    \label{eq.1}
\end{equation}

\textbf{Sensor emulation:} For bypassing the original restriction, the original \ac{EC} needs to be emulated with a manipulated \ac{TSS} signal. Therefore, a signal is to be emulated and amplified, depending on the current velocity (\textit{$v_{scooter}$}). The output frequency chosen is two times lower than the by the \ac{TSS} provided oscillating signal. That way, the \ac{EC} is not interfering the optimal ignition timing. For proper stimulation, the emulated signal has to be within the original speed-dependent voltage range ($\pm 30~V$). The emulated signal is generated by switching the 12 V battery voltage by a MOSFET, overcoming the \acp{EC} threshold. Equation (\ref{eq.2}) is used to calculate the emulating frequency (\textit{$f_{emul}$}) for the PWM timer, based on the second-stage transmission ratio (\textit{$i_{gear}$} = 13) and the number of poles (\textit{$n_{pole}$} = 4) attached to the clutch bell. For a velocity of max. 50 km/h, a frequency of 3.64 kHz results. To facilitate the comparison between the original and \ac{VC} system, a switch can be used to select between transferring the emulated or original \ac{TSS} signal to the \ac{EC}.

\begin{equation}
    f_{emul}(v_{scooter})=v_{scooter} \cdot \frac{2 \cdot i_{gear} \cdot n_{pole}}{D_{wheel}}
    \label{eq.2}
\end{equation}

\textbf{Fail-safe:} Every system interfering with vehicle dynamics, is to be classified as safety critical. The system manipulates longitudinal dynamics by throttle intervention. Depending on the error detected, the \ac{MECU} must react by predefined fail-safe states to ensure safe vehicle operation. Table \ref{Fail-safe} shows the error cases identified as possible. The listed fail-safe states are classified into four stages. Class 1 describes warnings that require re-calibration but do not danger safe operation. Class 2 can affect the operation by malfunction of the cruise control, which is why it is disabled. Class 3 can lead to total scooter misbehavior by loss of acceleration control. In this case, the \ac{TVA} is not accepting any throttle request. Class 4 represents the worst case. The \ac{MECU} can no longer control the powertrain and a contactor interrupts the ignition line to shut down the engine. A relay is to be integrated, controlling the ignition line. All error cases include monitoring of CAN bus communication, using safety features and life detection. 

\begin{table}[!h]
\footnotesize
\caption{List of fail-safe states}
\label{Fail-safe}
\begin{center}
\begin{tabular}{c c l c l}
\hline
\textbf{ID} & \textbf{C}                                   & \textbf{Error}          & \textbf{Type} & \textbf{Action}           \\ \hline
1           & \cellcolor[HTML]{FFFE65}1                        & \ac{TPS} calib. req. & Warning       & \ac{HMI} notification     \\ 
2           & \cellcolor[HTML]{FFFE65}1                        & \ac{TVA} calib. req. & Warning       & \ac{HMI} notification     \\ 
3           & \cellcolor[HTML]{FFCC67}{\color[HTML]{000000} 2} & \ac{HMI} CAN error   & Error         & Disable cruise contr.     \\ 
4           & \cellcolor[HTML]{F56B00}3                        & \ac{WSS} inaccuracy  & Error         & Reset \ac{TVA} \& set vel.\\
5           & \cellcolor[HTML]{F56B00}3                        & \ac{TPS} error det.  & Error         & Reset \ac{TVA} \& set vel.\\ 
6           & \cellcolor[HTML]{FE0000}4                        & \ac{TVA} error det.  & Error         & Engine shut down          \\ \hline
\end{tabular}
\end{center}
\end{table}

\subsection{Velocity control}
\begin{figure}[!h]
\centering
\includegraphics[width=8.5cm]{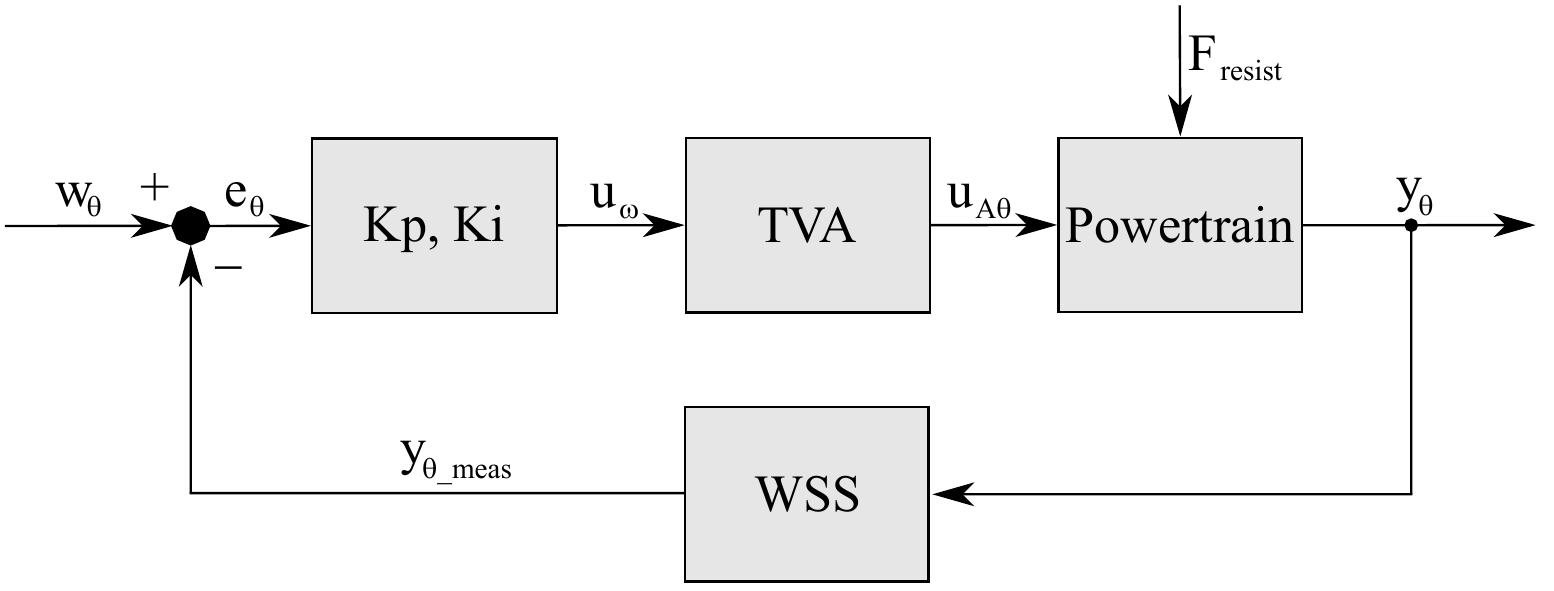}
\caption{\ac{VC} control loop}
\label{VC loop}
\end{figure}

Conventional vehicles require a throttle command by the rider that relates to the required power needed. The rider controls the desired vehicle velocity. The \ac{VC} approach describes a continuous control of the velocity. Instead of any required engine power, the throttle grip position represents a set velocity. Basically, the system performs a continuously varying cruise control algorithm. For this, a controller is designed, as shown in Figure \ref{VC loop}. The \ac{TPS} set velocity (w$_\theta$) serves as input reference variable. After determining the control deviation (e$_\theta$) by using the \ac{WSS} measurement as control variable feedback (y$_\theta$\textsubscript{\_meas}), the PI-controller outputs the manipulated variable u$_\omega$ in form of a throttle valve opening. Next, the \ac{TVA} controls the airflow (u$_{A\theta}$) within the air intake. Then, the powertrain reacts to the system's throttle demand and generates a propulsion force in relation to the throttle valve opening. Windy conditions, heavy load or street gradients can result in a disturbing/varying resistance force (F$_{resist}$). The vehicle velocity y$_\theta$ is measured by the \ac{WSS} and fed back.
  
\textbf{Modeling:} For adequate controller design, the system is simulated. Considering translatory ($F_{I_{trans}}$) and rotational ($F_{I_{rot}}$) mass inertia, air ($F_{Air}$) and rolling ($F_{Roll}$) resistance, the equilibrium of forces is described in (\ref{eq.3}).

\begin{equation}
    F_{Drive}= F_{I_{trans}} + F_{I_{rot}} + F_{Air} + F_{Roll} 
    \label{eq.3}
\end{equation}

In dependence on the vehicle position, a non-linear differential equation of second-order results for $F_{Drive}$ in (\ref{eq.4}).

\begin{equation}
\begin{split}
F_{Drive}(x)= &\ (m_{sc}+m_{r})\cdot\ddot{x}+\frac{(J_{f}+J_{r})}{r^2}\cdot \ddot{x} \\\
&+ \frac{A\cdot \rho \cdot c_{w}}{2}\cdot \dot x^2 + c_{r}\cdot (m_{sc}+m_{r})\cdot g
\end{split}
\label{eq.4}
\end{equation}

\begin{figure}[!h]
\centering
\includegraphics[width=8.8cm]{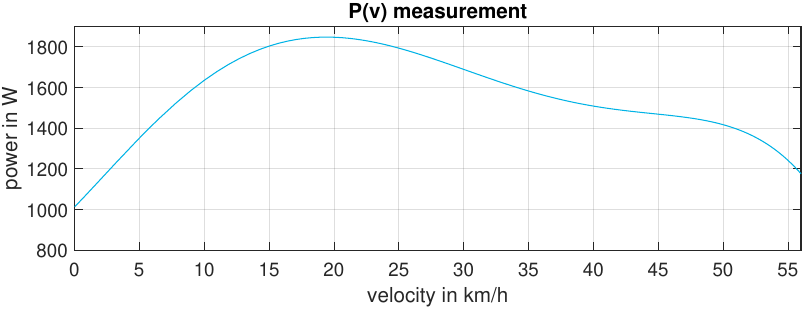}
\caption{Engine characteristics}
\label{Pv}
\end{figure}

$F_{drive}$ depends on the operating point of the engine and gearbox. To determine the power characteristics, performance measurements were taken on a roller dynamometer at various operating points, shown in Figure \ref{Pv}. With the clutch engagement, the power increases up to 15 km/h. The power declines with growing velocity due to the poor \ac{CVT} efficiency and the speed-dependent power drop of the engine. Caused by the test setup, the $F_{Roll}$ is part of the measured power and has to be compensated proportionally in case of partial load conditions. A polynomial ($P(v)_{poly}$) is derived from the data set. For a full throttle acceleration, the driving force can be determined by (\ref{eq.5}).   

\begin{equation}
    F_{Drive}(\dot{x})= \frac{P(v)_{poly}}{\dot{x}}
    \label{eq.5}
\end{equation}

The constants given in Equation (\ref{eq.4}) were measured or determined experimentally and are listed in Table \ref{Properties}.   

\begin{table}[!h]
\footnotesize
\caption{List of simulation related properties}
\label{Properties}
\begin{center}
\begin{tabular}{ c l c c c }
\hline
\textbf{}   & \textbf{Property}       & \textbf{Symbol}   & \textbf{Value}    \\ \hline
1           & Frontal area            & $A$               & 0.78 m$^2$        \\
2           & Wheel radius            & $r$               & 0.226 m           \\
3           & Inertia front wheel     & $J_f$             & 0.327 Nms$^2$     \\
4           & Inertia rear wheel      & $J_r$             & 1.3228 Nms$^2$    \\
5           & Mass scooter            & $m_{sc}$          & 99 kg             \\
6           & Mass rider              & $m_r$             & 80 kg             \\
7           & Air resistance coeff.   & $c_w$             & 0.64              \\
8           & Rolling resis. coeff.   & $c_r$             & 0.015             \\
9           & Air density             & $\rho$            & 1.225 kg/m$^3$    \\ \hline
\end{tabular}
\end{center}
\end{table}

\begin{figure}[!h]
\centering
\includegraphics[width=8.8cm]{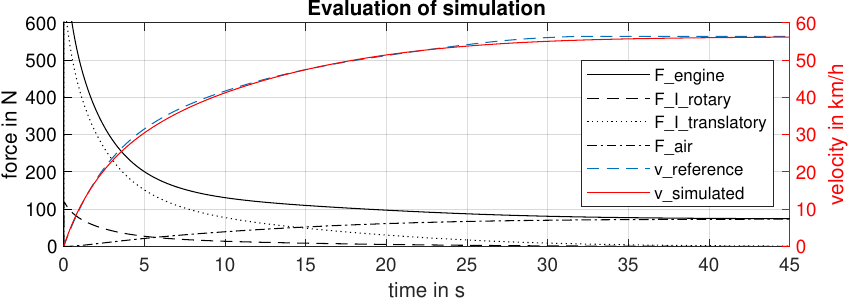}
\caption{Evaluation of simulation}
\label{SimEval}
\end{figure}

To check on model plausibility, the simulated step response was compared to a reference measurement, shown in Figure \ref{SimEval}. The final velocities and the acceleration behavior match. The maximum deviation is $\pm1$ km/h and the validity of the simulation is sufficient. In addition, the simulated forces are shown as further simulation output. 

\textbf{Control:} By applying the \ac{VC}, the throttle grip position equals a set velocity. Controlling the scooter's velocity constitutes a non-linear system. With increasing velocity, the control deviation rises due to higher driving resistance. Typically, control algorithms for non-linear systems are optimized for a certain operating point by linearization. Here, an adaptive PI-controller is proposed to change its characteristics velocity-dependent. This strategy achieves a balance between high controllability and control accuracy over the whole velocity range. In theory, a P-controller could already lead to adequate control behavior. If the gain is too high, not only the overshoot behavior but also signal disturbances of the velocity feedback would negatively affect the performance. Through adding an I-controller, the set velocity can be reached by smaller P-gains, stabilizing the controller and decreasing the susceptibility to disturbances. Therefore, the controller's parametrization needs to adapt continuously to the scooter's operating point. Proportional ($K_P$) and integral ($K_I$) gain adapts to the current scooter velocity ($v_{sc}$) given in (\ref{eq.6}) and (\ref{eq.7}). For both the max./min. gain is to be determined. In addition, a constant offset ($Os$) of the linear I gain curve prevents instability at very low velocities.
\begin{equation}
    K_{P}= \frac{v_{sc}}{v_{max}}\cdot (K_{P_{max}}-K_{P_{min})} + K_{P_{min}}
    \label{eq.6}
\end{equation}
\begin{equation}
    K_{I}= \frac{v_{sc}-Os}{v_{max}}\cdot (K_{I_{max}}-K_{I_{min})} + K_{I_{min}}
    \label{eq.7}
\end{equation}
Caused by the continuously changing set velocity and the varying steps within the throttle command, the I-controller can not perform precisely. Depending on the operating point, there are different settling times for different velocities and, as a result, varying integral values/overshoots. By implementing a set velocity-dependent threshold ($T_I$), the enabling time of the I-controller can be predefined related to the rising control deviation ($dev$) and set velocity ($v_{set}$). $T_I$ is described by (\ref{eq.8}). 
\begin{equation}
    T_{I} = \frac{v_{set}}{v_{max}}\cdot (K_{T_{max}}-K_{T_{min}}) + K_{T_{min}}
    \label{eq.8}
\end{equation}
If the control deviation overcomes the threshold in its current operating point, the integrator is reset by (\ref{eq.9}). The I-controller is only enabled in a certain error band in order to adjust the system as precisely as possible. This error band is centered around the set point and changes its size relative to the set point. This allows the control behavior to be kept consistent, regardless of the operating point and abrupt throttle commands. Additionally, the integrator is limited to prevent an integration wind-up when the actuator is saturated.
\begin{equation}
    \begin{split}
        (dev > T_I) \ \ \ I_{enbl} = 0 \ \ \ ; \ \ \  (dev \leq T_I) \ \ \ I_{enbl} = 1 
    \end{split}
    \label{eq.9}
\end{equation}
The throttle valve position ($Pos_{thr}$) is given by (\ref{eq.10}).
\begin{equation}
    Pos_{thr} = K_P(v_{sc})\cdot dev + K_I(v_{sc})\cdot I_{enbl}(T_I)\cdot \int dev
    \label{eq.10}
\end{equation}
Based on the longitudinal scooter model, a closed loop simulation was set up for tuning the controller. A max. overshoot of 0.3 km/h (barely noticeable for the rider) and robust control behavior, under consideration of sensor disturbances, were required. In addition to a fast settling time, driveability was also taken into account. At high gains, the powertrain reacts too sensitively to small throttle deflections, despite constantly stable control behavior. Therefore, a parameter set was chosen that combines the highest possible performance with enabling smooth accelerations. Using simulation, it was determined which parameter set was best with regard to control performance. This was then optimized for driveability in a road test. Table \ref{Ctr_Params} shows the resulting controller parameters. The algorithm is being executed with a cycle time of 20 ms.

\begin{table}[!h]
\footnotesize
\caption{List of controller parameters}
\label{Ctr_Params}
\begin{center}
\begin{tabular}{c c c c c c}
\hline
\textbf{}     & \textbf{$P$}  & \textbf{$I_{sim}$}  & \textbf{$I$}  & \textbf{$T_{I,sim}$}  &\textbf{$T_I$}  \\ \hline
$K_{min}$     & 2.1           & 0.02                & 0.01          & 1.44                  & 1.3            \\ 
$K_{max}$     & 16            & 0.07                & 0.075         & 6.84                  & 6.3            \\
$Os$          & /             & 3.0                 & 3.0           & 3.0                   & 3.0            \\ 
$Limit$ (\%)  & -100/100      & 0/100               & 0/100         & 0/100                 & 0/100          \\ \hline
\end{tabular}
\end{center}
\end{table}

\vspace{-0.3cm}

\subsection{\ac{HMI} - Virtual dashboard}
\begin{figure}[!h]
\centering
\includegraphics[width=7cm]{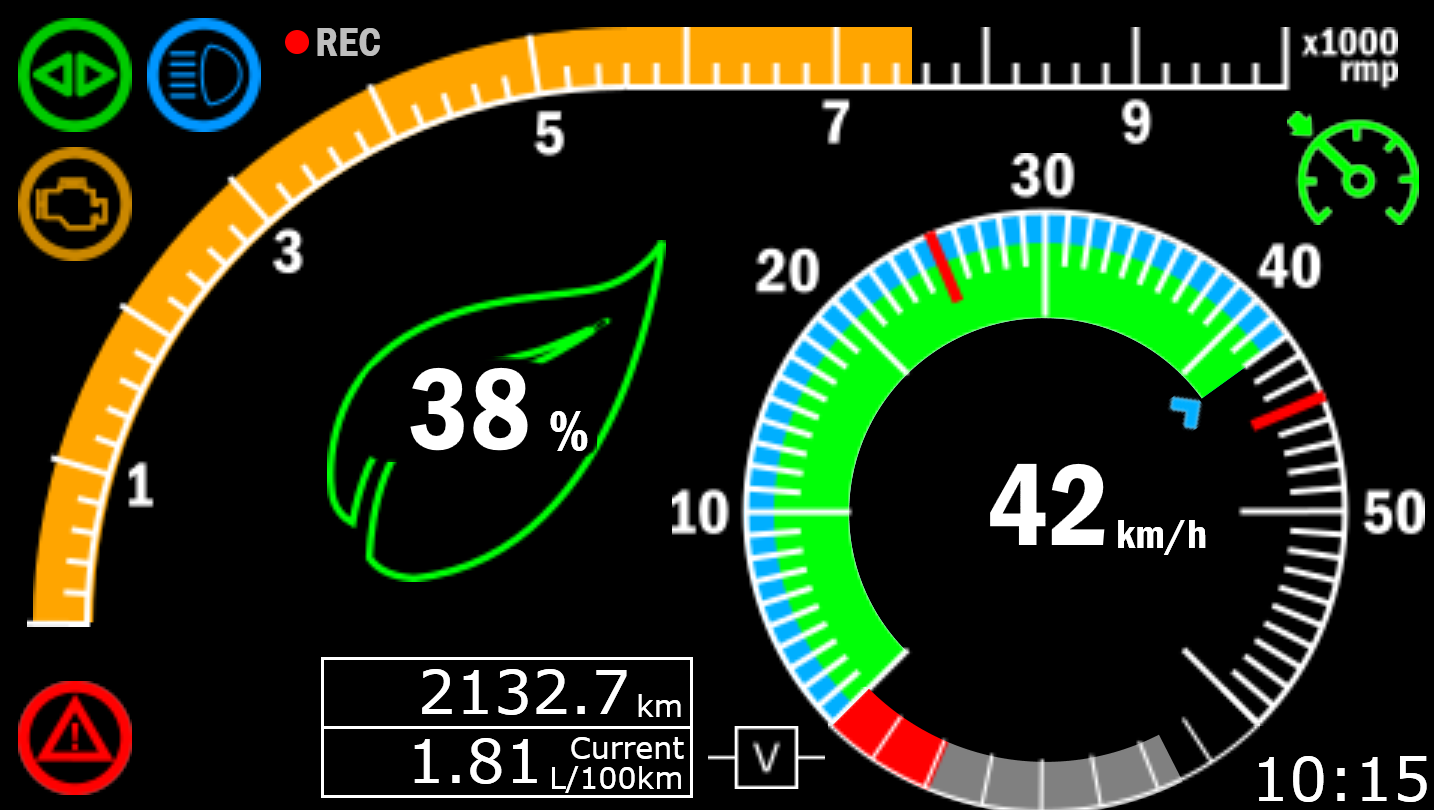}
\caption{\ac{HMI} virtual dashboard}
\label{HMI_fig}
\end{figure}

Virtual dashboards are uncommon in this vehicle class. In context of the \ac{VC}, cruise control and driving modes, they offer plenty of opportunities for adequate visualization and interaction. The \ac{HMI} consists of a 4.3" LCD display combined with a STM32 MCU (H7B3) \citep{STM32HMI}. Inputs were realized using a micro joystick integrated into the throttle armature. In addition to the original information, the following functions have been enhanced, as shown in Figure \ref{HMI_fig}.

\begin{enumerate}
    \item Set/scooter velocity: Based on the \ac{VC} strategy, the by the rider given velocity demand (blue) needs to be displayed in addition to the scooter velocity (green). 
    \item Cruise control interface: For enabling/disabling the cruise control, the selected velocity is set by a small arrow in combination with a cruise control sign.
    \item Eco-score: To challenge eco-friendly driving, an eco-score (green leaf) is built up with the fuel savings made.
    \item Evaluation information: The engine speed, amount of fuel injected and recording state of the \ac{MB} are to be displayed.
    \item Driving mode: By use of the \ac{HMI}, the developer can switch between the original restriction (\ac{TSS} signal/no \ac{VC}) and enabled \ac{VC}-System (\ac{TSS} emulated by \ac{MECU}). 
\end{enumerate}

\subsection{Measurement box}
For system integration, \ac{VC} parametrization and lastly functional verification, system states and dynamic vehicle data must be measured. Therefore, a \ac{MB} was designed for logging CAN bus data, measuring engine speed and injection cycle, storing files in real-time and offering data transfer via a mircoSD card. Set point and process value of the velocity as well as the throttle valve position are tapped directly from the vehicle's CAN bus. For the measurement of the current engine speed, the speed signal of the original crankshaft sensor was processed and evaluated. To measure vehicle dynamics, a 6-axis inertial measurement unit was used, fusing sensor values by complementary filters to measure orientation and acceleration. All measured quantities are periodically stored with a time stamp, provided by a real-time clock. As shown in Figure \ref{SysLayout}, the \ac{MB} (No. 5) is not part of the \ac{VC}-system. Starting or stopping a recording is done via the \ac{HMI}. New files are automatically created and adjustments to the recording rate or CAN message selection can be edited quickly. 

\section{Results} 

\subsection{Controller performance}
\begin{figure}[!h]
\centering
\includegraphics[width=8.7cm]{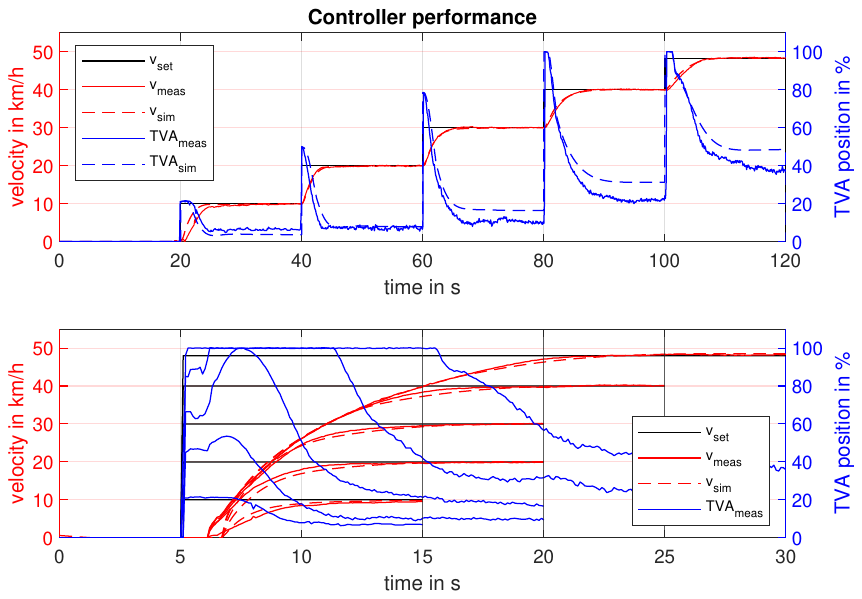}
\caption{Controller performance}
\label{CTR_Eval}
\end{figure}

To achieve stable, stationary and precise control, an adaptive PI-controller was tuned simulation-based (Table \ref{Ctr_Params}) under the criteria of max. 0.3 km/h overshoot. A test cycle, consisting of a series of increasing velocity steps (0, 10, 20,... km/h) and sudden accelerations from standstill (0, 10, 0, 20,... km/h), was performed. That way, the proper function of the adaptive threshold for larger steps in set point as well as small variations can be tested. By comparing the set velocity with the resulting simulated/measured velocity and the simulated closed-loop/measured \ac{TVA} position, the simulation validity and real-world control performance can be investigated. Figure \ref{CTR_Eval} presents the test results (top: incremental, bottom: initial). For both scenarios, the controller reached the exact set velocity without any overshoot or instability. 

\newpage \noindent
Velocities lower than 11 km/h led to unsteady settling due to the centrifugal clutch engagement/slip. For higher velocities, the max. error between simulated and measured dynamics is 1.9 \% for step-wise increments. Initial accelerations from standstill lead to shifted velocity origins that can be justified by the clutch engagement. The resulting error is max. 2.3 \% which proves a plausible controller simulation beyond the engagement process. Deviations in \ac{TVA} openings are due to the non-linear relationship between throttle opening and engine power output. This was neglected in the simulation, but has no measurable effect on the control behavior. Higher controller gains are feasible but lead to oversensitive response to throttle changes.

\subsection{System behavior}
\begin{figure}[!h]
\centering
\includegraphics[width=8.8cm]{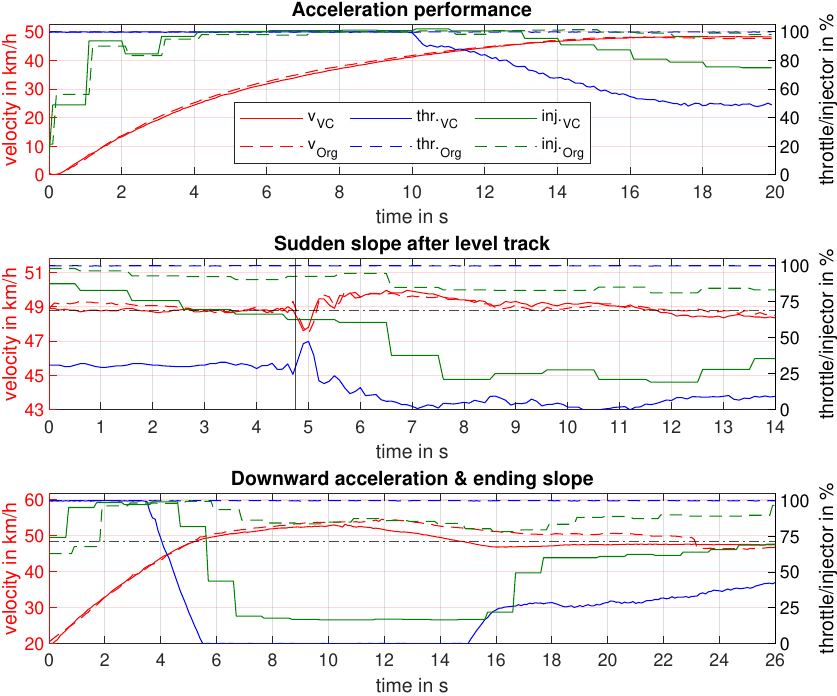}
\caption{System behavior}
\label{VC_behavior}
\end{figure}

After validation of the control strategy and performance, the effect on vehicle dynamics and engine control is to be examined in real-world scenarios. Three frequently occurring disturbance scenarios were analyzed: max. acceleration (S1), a sudden change from a level road to an uphill/downhill slope (S2) and an acceleration on a downward slope followed by a transition to a level (S3). The comparison to the original restriction is shown in Figure \ref{VC_behavior} by measurements of the velocity, throttle valve position and the fuel injected. Differences between scenarios result from small road gradient variations.

Scenario (S1) demonstrates the preservation of driving performance. Acceleration is equally fast with both systems, whereby the original system, contrary to the \ac{VC}, shows a slight overshoot of 0.3 km/h (\ac{VC}: 0 km/h) when reaching the speed limit. Significant differences can be found in the throttle valve position and injection quantity. When approaching the speed limit (orig. 48.7 km/h, dash-dotted line), the \ac{VC} regulates the throttle valve at the corresponding operating point (50 \%). Consequently, the injection quantity also drops by 25 \%. Scenario (S2) is initiated at top speed, which is why the \ac{VC} has already settled to approx. 35 \% throttle position. At time 4.75 s, a 8 \% downhill slope suddenly occurs after a small bump. Despite aggressive excitation, the system behavior remains stable and even resembles the response of the control applied by Peugeot. In both cases, the overshoot behavior at the entry of the gradient also shows max. 1 km/h. The \ac{VC} closes the throttle valve almost completely, whereas originally it's fully open and three times the fuel is injected. Scenario (S3) shows the max. acceleration on a 15 \% downhill slope followed by a transition to level ground. An improved overshoot (1.1 km/h smaller) can be seen with the \ac{VC}. While the speed limit can no longer be maintained in the original condition, it can be successfully controlled by the \ac{VC}. It is again shown that the throttle valve is completely closed and the injection is reduced to idle state. At time 16 s, the gradient ends. The system response is stable with no discernible response in velocity. 

\subsection{Fuel savings}
In order to demonstrate the fuel savings/CO$_2$ reduction under real-world conditions, a 50.4 km long test cycle was driven that depicts 61 \% urban and 39 \% rural environments. Three cycles each were driven for the original throttle and the \ac{VC}-system. Weather conditions, driving time and traffic situation were monitored to create comparable test conditions. The tire pressure was kept constant at the manufacturer's specification of 2.1 bar and the rider (85 kg) was not changed. Consumptions were determined by means of calibrated measuring cylinders. Inaccuracies of max. 2.5 ml occur due to measurement errors, which result in a max. total error of 0.3 \%. The measured volumes were referenced to a temperature of +15 °C. By applying the \ac{VC}, a saving of 13.6 \% could be demonstrated compared to the original condition (1.82/2.11 l/100 km). Since the combustion of one liter of gasoline releases 2280 g CO$_2$, this corresponds to an estimated reduction of approx. 661 g/100 km. Table \ref{T_fuelSaving} shows the related results.

\begin{table}[!h]
\scriptsize
\caption{List of Road testing results}
\label{T_fuelSaving}
\begin{tabular}{ccccccc}
\hline
\textbf{Mod.} & \textbf{Weath.} & \textbf{Dur.} & \textbf{Mile.} & \textbf{Vol.} & \textbf{Cons.} & \textbf{Aver.} \\
/             & °C               & h               & km              & ml           & l/100km            & l/100km        \\ \hline
Org.          & sun,dry,12       & 1:25            & 50.4            & 1060         & 2.10               &                \\
Org.          & sun,dry,16       & 1:26            & 50.4            & 1061         & 2.10               & \textbf{2.11}  \\
Org.          & sun,dry,17       & 1:31            & 50.4            & 1065         & 2.11               &                \\ \hline
VC            & sun,dry,17       & 1:29            & 50.4            & 925          & 1.83               &                \\
VC            & sun,dry,12       & 1:32            & 50.4            & 885          & 1.76               & \textbf{1.82}  \\
VC            & sun,dry,18       & 1:33            & 50.4            & 945          & 1.87               &                \\ \hline
\end{tabular}
\end{table}

This saving can be explained by the optimization of the ignition timing. On the flat, approx. 16 degrees after \ac{TDC} can be measured with original restriction. By applying the \ac{TbWS}, the air supply is controlled instead of the ignition timing. This allows the \ac{EC} to set the optimum ignition timing and reduce the injection quantity to maintain $\lambda$=1. Now the ignition timing corresponds to -5 degrees (before \ac{TDC}) at max. velocity and is thus in the optimum range (compare Fig. \ref{IgnT}). A significant improvement in exhaust gas composition can also be expected.

\section{Discussion}
Modern four-stroke scooters already enable a comparatively environmentally friendly type of individual mobility. The system developed increases efficiency by an additional 14 \%. More and more electric scooters are available as alternatives, but they come with significantly smaller ranges, heavy batteries and higher prices. If CO$_2$ emissions are considered in comparison, vehicle production and electricity generation must be considered. One liter of gasoline emits about 2280 g \citep{G_CO2} of CO$_2$ when burned, while the European average for the production of one kilowatt-hour of electrical energy is about 226 g of C0$_2$ \citep{E_CO2}. The average consumption of combustion-powered scooters from several leading manufacturers in the EU (Peugeot, Vespa, Piaggio) equals an average of 2.22 l/100 km (5143 gCO$_2$/100 km) for gasoline-powered and 4.11 kWh/100 km (928 gCO$_2$/100 km) for elec-tric-powered scooters. Taking into account the average battery capacity of the referred manufacturers, a capacity of 4.88 kWh results. The production emissions of both vehicle types differ mainly through battery production, which releases about 177 kg of CO$_2$ per kWh of capacity \citep{G_CO2}. Consequently, the electric drive generates an additional 864 kg offset of CO$_2$ on average. With a life expectancy of scooters/mopeds of 20000 km, the CO$_2$ emission for the gasoline-powered scooter is 1028.6 kg and 1049.6 kg for the electric-powered one. The test vehicle equipped with the velocity-controlled \ac{TbWS} achieves a CO$_2$ footprint of 829.9 kg during its service life with an average consumption of 1.81 l/100 km, making it 20.9 \% more environmentally friendly than average electric scooters. After 26800 km, the electric scooter would show a CO$_2$ benefit. When charging with electricity from renewable energy sources solely, the by battery production caused emissions would still overcome the test vehicle's footprint. By using eco-fuels, the CO$_2$ balance could even be neutral. Accordingly, consumption optimization would serve for economic efficiency and effective utilization of the eco-fuel resource \citep{E-fuel}.\\

Besides the eco-friendliness, the driving experience can be questioned. By \ac{VC} implementation, there is no direct command line to the engine. The system reacts proportionally to the control deviation, which is similar to conventional acceleration behavior. Like cruise control, the system can accelerate independently on uphill or downhill grades to maintain speed. The rider has to get used to this behavior, as it differs from conventional scooters. Other control approaches could be considered. Furthermore, the highest energy loss of 28 \% (test vehicle) occurs in the variomatic transmission. This design-related problem affects all scooters with \ac{CVT} belt drive \citep{CVT}. By developing a more efficient torque converter, the efficiency could be increased significantly. Consumption of less than 1.5 l/100 km could become feasible. Hydrodynamic torque converters could be investigated \citep{TorqueConv}.

\section{Conclusion}
In this article, the development of a velocity-controlled Throttle-by-Wire-System is presented. The objective, to provide a more eco-friendly restriction method, was successfully reached. Instead of shifting the ignition timing to comply with the legal maximum velocity for scooters, this system regulates the air supply to the engine. For this purpose, a wheel speed and throttle grip sensor were designed to sense feedback and control variable. A throttle actuator was designed to allow system intervention by engine power manipulation. Power supply, data processing, motor control emulation, fail-safe functions and actual control were implemented on a main \ac{ECU}. By simulating the vehicle's longitudinal dynamics, an adaptive controller could be designed to control the velocity stable, with stationary accuracy and without overshoot. The driveability/operability of the vehicle was prioritized. A virtual dashboard was used to interface between the system and the rider, clarifying the functionality and promoting eco-friendly driving with an eco-score. Additionally, it enables the activation and adjustment of the cruise control. In order to be able to evaluate the system behavior, a measurement box was integrated, logging all system parameters. Ultimately, a fuel saving of 13.6 \% was achieved and proven in the road test, while maintaining the vehicle's performance. As a result, the competitiveness against electrically powered scooters/mopeds was underlined. As long as large amounts of CO$_2$ are emitted during battery production, the optimized four-stroke powertrain presented here represents an excellent transitional solution. Mechanical optimizations could further reduce consumption. 

\section*{Acknowledgements}
The herein presented research was funded and supported by the University of Cadiz and Frankfurt University of Applied Sciences. The authors acknowledge the valuable support of Peugeot Motocycles Deutschland GmbH.


\bibliographystyle{elsarticle-harv} 
\bibliography{reference}



\end{document}